# Surface nanoscale axial photonics

M. Sumetsky and J. M. Fini

*OFS Laboratories, 19 Schoolhouse Road, Somerset, NJ  08873*

sumetski@ofsoptics.com

Dense photonic integration promises to revolutionize optical computing and communications. However, efforts towards this goal face unacceptable attenuation of light caused by surface roughness in microscopic devices. Here we address this problem by introducing Surface Nanoscale Axial Photonics (SNAP). The SNAP platform is based on whispering gallery modes circulating around the optical fiber *surface* and undergoing *slow axial propagation* readily described by the one-dimensional Schrödinger equation. These modes can be steered with dramatically small *nanoscale* variation of the fiber radius, which is quite simple to introduce in practice. The extremely low loss of SNAP devices is achieved due to the fantastically low surface roughness inherent in a drawn fiber surface. In excellent agreement with the developed theory, we experimentally demonstrate localization of light in quantum wells, halting light by a point source, tunneling through potential barriers, dark states, etc. This demonstration, prototyping basic quantum mechanical phenomena with light, has intriguing potential applications in filtering, switching, slowing light, and sensing.

For the past few decades, research and development in photonic integrated circuits [1-9] has struggled to build a platform with the miniature dimensions, flexibility, and control needed to deliver breakthrough capabilities in optical computing, communications, and fundamental science. The platform must include waveguides for propagation and routing light, but must also provide means for trapping light. Building microscopic high-quality-factor traps for photons is non-trivial. One cannot simply construct a microscopic version of a traditional optical resonator, since naturally reflecting materials (metals) are highly absorbing. Also, one cannot construct photonic quantum wells using a straightforward analogy to electronics and quantum mechanics. Indeed, this would require total internal reflection of light everywhere at the boundary surrounding the trap, a challenge for the low-loss optical cavities with microscopic dimensions [6].

Two basic platforms for fabrication of photonic circuits with microscale elements have been developed. The first, a ring resonator platform, is built on thin transparent photonic wires that tightly confine light due to the high refractive index contrast between the wire and surrounding material. In this platform, wires form closed rings that are coupled together to provide higher level of functionality [1,2,8,9]. The second platform is based on photonic crystals, materials with periodic modulation of refractive index [5]. Thin waveguides are created using a linear disruption of periodicity while microresonators are formed by localized defects [3-7]. With the advances in lithographic technology and design methods, these two platforms have steadily increased in complexity, achieved lowed loss, and broadened functional capability. The miniature integrated photonic circuits that have been proposed and demonstrated using the ring-resonator and photonic crystals platforms can potentially revolutionize optical communications and computing by implementation of superfast all-optical processors of light pulses on a chip and, in particular, miniature optical buffers [1,3] and all-optical switches [2,6].

Yet, despite the remarkable accomplishments, the existing platforms still face a severe limitation: It is necessary but very difficult to fabricate photonic circuits that have both *microscopic dimensions* and *ultra-small losses* (required, e.g., for creating the ultra-high quality-factor microresonators). Indeed, the microscopic dimensions of photonic elements can only be achieved with high-index-contrast material interfaces. However, precisely in this regime of high index contrast, the sensitivity to surface roughness and fabrication errors is daunting and potentially fundamentally limiting. To reduce the sensitivity to the interface roughness and uncontrolled attenuation, photonic circuits with low-index contrast can be exploited [9,15]. However, to avoid radiation losses at low-contrast waveguide bends, the size of photonic resonators has to have the millimeter scale [9,10]. These photonic circuits might be low-loss and easier to fabricate, but they are no longer microscopic.

In this Article, we present a new photonic platform with ultra-low loss, flexibility, and elements having microscopic dimensions. Our solution consists in Surface Nanoscale Axial Photonics (SNAP) employing whispering gallery modes (WGMs) [11-14] that circulate circumferentially around the surface of a thin optical fiber while also undergoing slow propagation along the fiber axis. SNAP elegantly manifests a number of key properties:

(a) *Remarkably low surface roughness* effortlessly results from the fiber draw process. The key benefit of SNAP compared to lithography-based technologies is the *orders of magnitude lower loss* achieved at a high-index-contrast silica surface.
(b) *Periodicity and slow light without the periodic modulation of the refractive index* (as for photonic crystals): Periodicity is introduced automatically by each revolution around the fiber surface. Axial propagation naturally has features of slow light, since the propagation of slow WGMs is primarily azimuthal.
(c) *SNAP is readily described by the one dimensional Schrödinger equation:* The axial propagation of slow WGMs exhibits turning points, localization in quantum wells, tunneling through barriers, etc.
(d) *Precise control of light achieved with nanoscale variation of the effective fiber radius,* which define the "potential" that steers, localizes, and engineers WGMs.
(e) *Novel phenomenon of surface WGMs* completely halted with a *point* source of light.
(f) *Microscopic length scale in each dimension:* For slow WGMs, the characteristic *axial wavelength is much larger than the wavelength of light* and has the order of a few tens of microns. This dramatically simplifies fabrication of SNAP devices and, in particular, the ring microresonators. As opposed to the macroscopic increase of the diameter of ring resonators in low-contrast photonic circuits, we increase the third dimension along the fiber axis and keep it microscopic.

**SNAP theory: prototyping one-dimensional quantum mechanics with light**

Modes in an optical fiber are characterized by the propagation constant $\beta(\lambda,z)$ which depends on the radiation wavelength $\lambda$ and variations of the fiber radius, $r(z)=r_0+\Delta r(z)$ and refractive index $n_f(z)=n_{f0}+\Delta n_f(z)$. In conventional optical fibers, light is directed along the interior fiber core and has the propagation constant close to $\beta_0(\lambda)=2\pi n_{f0}/\lambda$. In contrast, SNAP employs transverse WGMs wrapped around the fiber surface by total internal reflection. The propagation constant of these modes is much smaller than $\beta_0(\lambda)$ and the speed of their axial propagation is much smaller than the speed of light in the fiber material, $c/n_{f0}$. In fact, the axial speed of a WGM and its propagation constant can be zero at the resonance wavelength $\lambda_{res}$ defined by the condition of *stopped axial propagation* $\beta(\lambda_{res}+i\gamma_{res},z)=0$, where the resonance width $\gamma_{res}$ determines the propagation loss.

The central idea of SNAP is to exploit the sensitivity of WGMs to the extremely small variations of the fiber radius and index near the resonance $\lambda_{res}$. Generally, a variation in radius causes coupling between modes and intermodal transitions, a complex problem which is usually addressed with the system of coupled wave equations [15]. In SNAP, this problem is absent: Variation $\Delta r(z)$ and $\Delta n_f(z)$ is so small and smooth that the coupled wave equations become decoupled and a single WGM is defined by a single differential equation. Near the resonance wavelength $\lambda_{res}$, this equation takes the form of the stationary Schrödinger equation (see Supplementary Information, Section 1):

$$\frac{d^2 A}{dz^2}+\beta^2(\lambda,z)A=0, \quad \beta^2(\lambda,z)=E(\lambda)-V(z),$$
$$E(\lambda)=-2\beta_0^2(\lambda_{res})\frac{\lambda-\lambda_{res}-i\gamma_{res}}{\lambda_{res}}, \quad V(z)=-2\beta_0^2(\lambda_{res})\left(\frac{\Delta r(z)}{r_0}+\frac{\Delta n_f(z)}{n_{f0}}\right),$$
(1)

where the effective energy $E(\lambda)$ is proportional to the wavelength variation and the effective potential $V(z)$ is proportional to the radius and index variation. Introducing the effective (optical) radius $r_{eff}(z)=n_f(z)r(z)$, we can simply say that the potential in Eq. (1) is proportional to the effective radius variation, $\Delta r_{eff}(z)=r_0\Delta n_f(z)+n_{f0}\Delta r(z)$.

Near the resonance wavelength $\lambda_{res}$, dramatically small effective radius variations are enough to reflect and confine light. Indeed, it follows from Eq. (1) that for the typical experimentally determined value of $\gamma_{res} \sim 10^{-7}$ μm at wavelength $\lambda_{res} \sim 1$ μm and fiber radius $r_0 \sim 10$ μm, light can be confined with an effective radius variation of the order of $r_0\gamma_{res}/\lambda_{res} \sim 1$ pm. This corresponds to feature perturbations of $\Delta r(z) \sim 1$ pm and $\Delta n_f(z) \sim 10^{-7}$. The characteristic axial wavelength in this case is $2\pi/\beta(\lambda,z) \sim \lambda_{res}^{3/2}\gamma_{res}^{-1/2}n_f^{-1} \sim 1$ mm, i.e., 1000 times greater than the wavelength of light.

Slow WGMs can be excited in an optical fiber using a microfiber (MF) [16], specifically, a micrometer diameter waist of a biconical fiber taper, which is attached normal to the SNAP device and connected to the light source and detector as illustrated in Fig. 1. The MF waveguide acts as a point source in Eq. (1) so that the amplitude of the WGM excited by the MF positioned at $z=z_1$ is expressed through the Green's function $G(\lambda,z_1,z)$ of Eq. (1),

$$\Lambda(\lambda,z_1,z)=CG(\lambda,z_1,z).$$
(2)

The spectrum of excited WGMs can be observed by measuring the transmission amplitude of the MF,

$$T(\lambda,z_1)=1-i|C|^2 G(\lambda,z_1,z_1). \qquad (3)$$

Parameter $C$ in Eq. (2) and (3) determines coupling to the MF. It is a weak function of wavelength and, with good accuracy, is constant in the vicinity of resonance $\lambda_{res}$. The derivation of Eq. (2) and Eq. (3) based on the formalism of the Lippmann-Schwinger equation [17,18] is given in Section 2 of the Supplementary Information.

SNAP devices include the three basic building blocks illustrated in Fig. 2. The first is the WGM bottle microresonator [19] coupled to a MF (Fig. 2(a)). This corresponds to a quantum well $V(z)$ in Eq. (1), and leads to the formation of states localized between turning points $z_{t1}$ and $z_{t2}$ at a discrete series of wavelengths corresponding to resonances in the transmission amplitude $T(\lambda,z_1)$. From Eq. (3), the resonance transmission amplitude $T(\lambda,z_1)$ is proportional to the WGM amplitude at the MF position $z=z_1$. In particular, if $z_1$ coincides with a node of the WGM then the coupling to the MF vanishes and the WGM becomes dark, i.e., it practically does not show up in $T(\lambda,z_1)$ (for details, see Supplementary Information, Section 3.3).

A second building block is a concave fiber waist (Fig. 2(b)). In this case, for wavelengths in the underbarrier region, $E<V(z)$, the WGM is localized due to the exponential decay of its amplitude away from the MF position $z_1$. Alternatively, above the barrier, for $E>V(z)$, the excited WGM is delocalized. In the particular case of a uniform fiber, $\Delta r_{eff}(z)\equiv 0$, Fig. 2(b) offers a simple explanation of the localization of light in a uniform cylindrical microresonator described with a less general semiclassical theory in [20] (see Supplementary Information, Section 3.1).

In the third building block (Fig. 2(c)), the MF is positioned near a turning point $z_t$ where the SNAP device has monotonically increasing radius. In this case, the wave that is launched by the MF along the positive axial direction, interferes with the wave that is launched along the negative direction and reflects from the turning point $z_t$. At discrete wavelengths when the *condition of destructive interference* of these two waves is fulfilled, the distribution of light is fully localized between the turning point $z_t$ and MF position $z_1$. Fig. 2(c) offers a simple explanation of the remarkable effect when *a point contact with a MF source completely halts light propagating along the SNAP device*, while the detailed theory of this effect is given in Section 3.2 of the Supplementary Information. In particular, Fig. 2(c) clarifies the appearance of localized conical modes discovered in [21].

Generally, a SNAP device includes a series of these building block elements, which is coupled to one or more transverse MFs. The field distribution along the SNAP device as well as the transmission spectrum, group delay, and dispersion of light transmitted through MFs can be engineered using Eq. (1), (2), and (3). In the experiments below we consider a SNAP device which reproduces the nanoscale radius variation of Fig. 2. In doing this, we experimentally demonstrate all the described basic elements and find excellent agreement of their performance with the developed theory.

**Experiment: Localization in quantum wells, tunneling, dark states**

The described basic SNAP phenomena, familiar from elementary quantum mechanics [22], are confirmed experimentally in excellent agreement with the developed theory. In our experiments, to arrive at small dimensions of SNAP devices, we fabricated them from a regular optical fiber drawn down to $r_0 \approx 11 \mu m$. The nanoscale variation of the fiber radius, $\Delta r(z)$, was introduced using a simple technique of controlled local heating with $CO_2$ laser and pulling. Thus we avoid the surface roughness that might result from other methods of fiber post-processing. Other methods include annealing, laser polishing, and UV exposure of photosensitive fibers. Due to the drastic elongation along the fiber axis, the field distribution in SNAP devices is much easier to access, control, and engineer as compared to the WGM-based microdevices demonstrated previously including spherical, toroidal, and ring microresonators [13,14] as well as microresonators directly fabricated by relatively small deformation of the optical fiber surface [23].

We fabricated samples of SNAP devices reproducing the characteristic *nanoscale* fiber radius variation illustrated in Fig. 2, the shape of *an elongated bottle (quantum well) with a neck (barrier)*. These samples include bottle microresonators with multiple axial localized states (Fig. 3(b), (c)), three axial localized states (Fig. 3(d), (e)) and a single axial localized state (Fig. 3(f), (g)). The samples were experimentally characterized as follows. First, an MF was translated along the test fiber in 20 μm steps where the transmission amplitude spectra (vertical plots in Fig. 3(b), (d), and (f)) were measured with an optical vector analyzer (1.3 pm wavelength resolution). The fiber radius variations (bold curves in Fig. 3(c), (e), and (g)) are found from these plots by applying the MF scanning method [24,25] corrected with Eqs. (1) and (3) (see Supplementary Information, Section 5).

In excellent agreement with theory, we observed full localization of light in dramatically shallow bottle microresonators. In Fig. 3(b), (c), the fiber shape features an elongated (300 μm in axial length) and extremely shallow (only 7 nm in radius variation) bottle microcavity. In agreement with the developed theory, the bottle resonances in Fig. 3(b) are localized between turning points $z_{t1}$ and $z_{t2}$. Depending on the value of coupling with the MF, the Q-factor of these resonances varies from a relatively low ($Q \sim 3 \cdot 10^4$) or greater than $10^6$ (with the resolution limited by the measurement device). The resonance state can be dark (i.e., practically uncoupled from the MF) if the MF position approaches a node of the localized state or is deep in the underbarrier region (see Supplementary Information, Section 3.3). Approximating the radius variation in Fig. 3(b) and (d), with the quadratic dependence, $\Delta r(z) = -z^2 / 2R$, we find the axial radius-of-curvature $R = 0.93$ m of the multi-level bottle microresonator (Fig. 3(c)) and a smaller $R = 0.04$ m for the three-level bottle microresonator (Fig. 3(e)). The experimental free spectral range $\Delta \lambda_{FSR}$ of the resonances in Fig. 3(b) and (d) is 0.082 nm and 0.38 nm, which is in good agreement with the theoretical values 0.08 nm and 0.39 nm found from Eq. (1) for the harmonic oscillator, $\Delta \lambda_{FSR} = \lambda_{res}^2 (2\pi n_f)^{-1} (r_0 R)^{-1/2}$ (see Supplementary Information, Section 4). Finally, Fig. 3(f) and (g) demonstrate a shallow bottle resonator with a single axial state. The length of this resonator is 70 μm and its height is only 5 angstrom. It is remarkable that such a minor deviation from uniformity is able to fully confine light.

Oscillations of spectra outside the quantum wells in Fig. 3 are easily explained. These oscillations result from the interference between light propagating directly through the MF and light, which couples into the test fiber, reflects from a turning point, and then couples back into the MF. Interestingly, on the right hand side of the barriers in Figs. 3(b), (d), and (f) there are high Q-factor resonances, which coincide with the quantum well resonances but are situated outside of the quantum wells. Appearance of these resonances is explained with Fig. 2: Light excited by a MF in the region of Fig. 2(c) propagates in direction of the quantum well, tunnels through the barrier, and resonantly attenuates. Finally, if the MF is positioned deeply in the underbarrier region (bottom of Figs 3(b), (d), and (f)), the outgoing WGMs strongly decay, interference is absent, and the transmission spectra are smooth.

**Experiment: Halting light with a point source**

We have experimentally demonstrated the remarkable effect of halting light with a point source, as predicted from our theory (see Fig. 2(c) and the description above). This effect takes place takes for the SNAP device illustrated in Fig. 4(a) when light excited at $z_1$ reflects from $z_t$ and destructively interferes at $z_1$. The transmission spectra and radius variation of the device, measured as in the previous case, are shown in Fig. 4(b) and (c). The localization of light is confirmed experimentally with two microfibers, MF1 and MF2. MF2 is translated along the test fiber and probes the field excited by MF1 as illustrated in Fig. 4(a). The measurement results are shown in Fig. 4(d), where the vertical axis lines of all spectra correspond to zero transmission. It is seen that at the discrete values of wavelengths indicated by horizontal arrows, the field distribution is fully localized along the fiber segments with a length of less than 100 µm and a radius variation of less than 5 nm. To compare the measurement results with theory, the fiber radius variation is approximated by the quadratic dependence $\Delta r(z) = z^2/2R$, and the Green's function of Eq. (1) is found analytically [26] (see Supplementary Information, Section 4). The axial fiber radius, $R=3.1$ m, is found from Fig. 4(c). With this value, the comparison of the transmission spectra measured experimentally and calculated with Eq. (3) and Eq. (A4.2) of the Appendix (respectively, black and blue curves in Fig. 4(b)) shows excellent agreement. A minor deviation near the principal peak maxima is explained by the deviation of the actual radius variation from the quadratic dependence away from its minimum. Finally, for comparison, the surface plot of the theoretical calculation of the WGM field amplitude distribution found from Eq. (A4.2) for the parameters of the experiment, Fig. 4(f), is placed in the background of Fig. 4(d). Good agreement is found both for the positions of the localized states and for the field distribution. A deviation near the principal peaks is, again, due to the assumed quadratic approximation for the radius variation.

**Discussion**

Nanoscale variation of the effective optical fiber radius (including the variation of the fiber radius and/or refractive index) enables formation and integration of independent or coupled microdevices (e.g., a sequence of microresonators) on a SNAP platform. The axial radiation wavelength of SNAP microdevices is significantly larger than the wavelength of light, which simplifies their operation and broadens the field of potential applications. The further decrease of the axial size of SNAP elements can be achieved with larger variation of the effective fiber radius while retaining adiabatic behavior and adherence to the developed theory. Due to the very low attenuation of light propagation along the optical fiber surface, the SNAP microdevices can exhibit significantly improved performance in filtering, time delay, slowing light, switching, sensing, etc., compared to lithographically fabricated photonic circuits.

**Acknowledgements**

The authors are grateful to Y. Dulashko for assisting in the experiments, to D. J. DiGiovanni and M. Fishteyn for useful discussions and suggestions.

**Supplementary Information**

**1. Derivation of Eq. (2) for the propagation constant of a slow WGM**

The propagation constant $\beta$ of optical modes in a fiber with radius $r_0$ and refractive index $n_f$, which is situated in the surrounding medium with index $n_0$, is defined by the equation [15]

$$(F_{1m}(U)+F_{2m}(W))\left(F_{1m}(U)+\frac{n_0^2}{n_f^2}F_{2m}(W)\right)=\left(\frac{m\beta}{kn_f}\right)^2\left(\frac{V}{UW}\right)^4, \quad (A1.1)$$

$$F_{1m}(x)=\frac{1}{xJ_m(x)}\frac{dJ_m(x)}{dx}, \quad F_{2m}(x)=\frac{1}{xH_m^{(2)}(x)}\frac{dH_m^{(2)}(x)}{dx} \quad (A1.2)$$

$$U=r_0(k^2n_f^2-\beta^2)^{1/2}, \quad W=r_0(k^2n_0^2-\beta^2)^{1/2}, \quad V=kr_0(n_f^2-n_0^2)^{1/2}, \quad k=2\pi/\lambda. \quad (A1.3)$$

where $J_m(x)$ and $H_m^{(2)}(x)$ are the Bessel and Hankel function, respectively. In the zero approximation, the propagation constant is set to zero, $\beta=0$, and Eq. (A1.1) is split into two equations for TE and TM modes:

$$F_{1m}(U_0)+F_{2m}(W_0)=0 \quad \text{(TE modes)} \quad (A1.4)$$

$$F_{1m}(U_0)++\frac{n_0^2}{n_f^2}F_{2m}(W_0)=0 \quad \text{(TM modes)} \quad (A1.5)$$

$$U_0=n_f kr_0, \quad W_0=n_0 kr_0. \quad (A1.6)$$

In the semiclassical approximation, for the WGMs located close to the fiber surface, $U_0\approx m\gg 1$, the resonant wavelengths are approximately defined from Eq. (A1.3) and (A1.4) as

$$\lambda_{mp}^{(0),\pm}\approx\lambda_m^{(0)}\left[1+\frac{\zeta_p}{2^{1/3}m^{2/3}}+\frac{n_0}{m(n_f^2-n_0^2)^{1/2}}\left(\frac{n_f}{n_0}\right)^{\pm 1}\right], \quad \lambda_m^{(0)}=\frac{2\pi n_f r_0}{m}, \quad (A1.7)$$

where $\zeta_p$ is a root of the Airy function ($\zeta_0\approx 2.338$, $\zeta_1\approx 4.088$, $\zeta_0\approx 5.22$), $p\sim 1$, and signs + and – correspond to TE and TM polarizations. From Eq. (A1.7), the free spectral range (FSR) is

$$\lambda_m^{(0)} = \frac{(\lambda_m^{(0)})^2}{2\pi n_f r_0}, \tag{A1.8}$$

Assume now that the wavelength $\lambda$, the fiber radius $r$, and the fiber refractive index are slightly shifted from $\lambda_{mp}^{(0)}$ and $r_0$, respectively. For a WGM propagating close to the fiber surface, in the first approximation in $\Delta\lambda = \lambda - \lambda_{mp}^{(0)}$ and $\Delta r = r - r_0$, iteration of Eq. (A1.1) near the resonance $\lambda_{res} = \lambda_{mp}^{(0)}$ yields

$$\beta^2 = 2\left(\frac{2\pi n_f}{\lambda_{res}}\right)^2 \left[\frac{\Delta r}{r_0} + \frac{\Delta n_f}{n_{f0}} - \frac{\Delta\lambda}{\lambda_{res}}\right]. \tag{A1.9}$$

This equation coincided with the expression for $\beta(\lambda,z)$ in Eq. (1).

## 2. Derivation of Eq. (2) for the WGM field distribution and Eq. (3) for the resonant transmission amplitude

Generally, the field excited by a MF in the SNAP device is localized in the vicinity of the MF/SNAP fiber contact point and does not have the rotational symmetry. However, near the resonance, $\lambda = \lambda_{res}$, the beam launched by the MF constructively interferes in the process of circulation along the SNAP fiber surface. Then, after a large number of turns, the beam acquires axial symmetry. The axially symmetric component of the beam becomes much greater in amplitude compared to its original asymmetric part, so that the latter can be ignored.

For the nanoscale and adiabatic variation of the SNAP fiber radius and outside the region of coupling with the MF, i.e., in the absence of source, Eq. (1) is the known uncoupled wave equation [15] with the propagation constant defined by Eqs. (A1.9). For a MF with the diameter ~ 1 μm, the axial size of the coupling region is $\Delta z_c \sim 1$ μm. Near the resonance $\lambda \approx \lambda_{res}$, $\Delta z_c$ is much smaller than the axial wavelength $\lambda_z = 2\pi/|\beta|$. Indeed, from Eq. (A1.9), for $\lambda \sim 1$ μm, $r_0 \sim 10$ μm, $\Delta r < 10$ nm, and $|\lambda - \lambda_{res}| < 1$ nm, we have $\lambda_z \geq 20$ μm. Thus, the effect of the MF positioned at $z = z_1$ can be described by adding a δ-function point source $C\delta(z - z_1)$ to the right hand side of Eq. (1). This immediately yields the expression for WGM amplitude defined by Eq. (2). More accurately, Eq. (2) for the WGM distribution and Eq. (3) for the transmission amplitude can be derived based on the formalism of the Lippmann-Schwinger equation. Following [17], the Hamiltonian describing the electromagnetic waves in the optical fiber coupled to a MF waveguide is approximated by $\mathbf{H} = \mathbf{H}_0 + \mathbf{V}$, where $\mathbf{H}_0 = \sum_{\mathbf{k}} E_{\mathbf{k}} |\mathbf{k}\rangle\langle\mathbf{k}| + \sum_{\mathbf{c}} E_{\mathbf{c}} |\mathbf{c}\rangle\langle\mathbf{c}|$ is the Hamiltonian of the uncoupled states in the MF and SNAP fiber and $\mathbf{V} = \sum_{\mathbf{c}_1 \neq \mathbf{c}_2} V_{\mathbf{c}_1,\mathbf{c}_2} |\mathbf{c}_1\rangle\langle\mathbf{c}_2| + \sum_{\mathbf{c},\mathbf{k}} \left(V_{\mathbf{c},\mathbf{k}} |\mathbf{c}_1\rangle\langle\mathbf{k}| + V_{\mathbf{k},\mathbf{c}} |\mathbf{k}\rangle\langle\mathbf{c}_1|\right)$ defines the MF/SNAP fiber coupling. Here $|\mathbf{k}\rangle$ is an uncoupled waveguide mode of the single mode MF and $|\mathbf{c}\rangle$ is an uncoupled WGM. For the axially symmetric fiber with very small radius variation, in the adiabatic approximation, the WGM

$\Psi_c(\mathbf{r}) = \langle \mathbf{r}|\mathbf{c}\rangle$ factors in the cylindrical coordinates $\mathbf{r}=(z,\rho,\varphi)$ into a product of axial $A_{m,p,q}(z)$, radial, $\Xi_{m,p}(\rho)$, and azimuthal, $\Phi_m(\varphi)$, components:

$$\Psi_c(\mathbf{r}) = A_{m,p,q}(z)\Xi_{m,p}(\rho)\Phi_m(\varphi) \quad (A2.1)$$

$m$ is the discrete azimuthal quantum number, $p$ is the discrete radial quantum number, $q$ is the discrete or continuous axial quantum number, and the amplitude $A_{m,p,q}(z)$ satisfied Eq. (1). All the factors in Eq. (A2.1) are assumed to be normalized (e.g., $\Phi_m(\varphi)=(2\pi)^{-1}\exp(im\varphi)$). The coupling operator $\mathbf{V}$ is spatially localized in a small vicinity of the MF contact point with the coordinate $z=z_1$. As mentioned, the axial length of this region is much smaller than the characteristic variation length $\lambda_z$ of $A_{m,p,q}(z)$ so that the coupling elements of this operator, $V_{\mathbf{c},\mathbf{k}} = \langle \mathbf{c}|V|\mathbf{k}\rangle$, are recast as

$$V_{\mathbf{c},\mathbf{k}} = C_{m,p}A_{m,p,q}(z_1), \quad C_{m,p} = \langle \Xi_{m,p}\Phi_m|V|\mathbf{k}\rangle. \quad (A2.2)$$

The total wave function is determined by the Lippmann-Schwinger equation,

$$|\Psi_{tot}\rangle = |\mathbf{k}\rangle + (E_\mathbf{k} - \mathbf{H}_0 + i\varepsilon)^{-1}\mathbf{V}|\Psi_{tot}\rangle = \mathbf{T}|\mathbf{k}\rangle, \quad (A2.3)$$

where $\varepsilon$ is a positive infinitesimal number enforcing the outgoing boundary condition and $\mathbf{T}$ is the scattering $\mathbf{T}$-matrix. After approximate diagonalization and renormalization, the elements of the $\mathbf{T}$-matrix, which determine the transition from the incoming wave $|\mathbf{k}\rangle$ into the WGM $|\mathbf{c}\rangle$ are [17,18]

$$T_{\mathbf{c},\mathbf{k}} = \frac{V_{\mathbf{c},\mathbf{k}}}{E_\mathbf{k} - E_\mathbf{c} + i\Gamma_\mathbf{c}} \quad (A2.4)$$

where $\Gamma_\mathbf{c}$ is the full resonant width of the state $|\mathbf{c}\rangle$, which takes into account both transmission losses in the SNAP fiber and leakage in the MF/SNAP fiber coupling region. The part of $\Psi_{tot}(\mathbf{r})$ localized in the SNAP fiber which is excited by the income wave $|\mathbf{k}\rangle$ is

$$\Psi_{TF}(\mathbf{r}) = \sum_\mathbf{c} \langle \Psi_{tot}|\mathbf{c}\rangle \Psi_\mathbf{c}(\mathbf{r}) = \sum_\mathbf{c} T_{\mathbf{c},\mathbf{k}}\Psi_\mathbf{c}(\mathbf{r}) \quad (A2.5)$$

Taking into account Eqs. (A2.3), (A2.4), and (A2.5), we find:

$$\begin{aligned}\Psi_{TF}(\mathbf{r}) &= \sum_\mathbf{c} \langle \Psi_{tot}|\mathbf{c}\rangle \Psi_\mathbf{c}(\mathbf{r}) \\ &= \sum_{m,p,q} \frac{C_{m,p}A_{m,p,q}(z_1)A_{m,p,q}(z)}{E_\mathbf{k} - E_\mathbf{c} + i\Gamma_\mathbf{c}}\Xi_{m,p}(\rho)\Phi_m(\varphi) \\ &= \sum_{m,p} C_{m,p}G_{m,p}(\lambda,z,z_1)\Xi_{m,p}(\rho)\Phi_m(\varphi),\end{aligned} \quad (A2.6)$$

where $G_{m,p}(\lambda,z,z_1)$ is the Green's function of Eq. (1) at $\lambda_{res}=\lambda_{mp}^{(0),\pm}$. In the vicinity of the resonance $\lambda \approx \lambda_{m_0 p_0}^{(0),\pm}$ all the terms in the second line of this equation except for those with $(m,p)=(m_0,p_0)$ can be ignored so that

$$\Psi_{TF}(\mathbf{r})\big|_{\lambda \approx \lambda_{mp}} \approx C_{m,p} G_{m,p}(\lambda,z,z_1) \Xi_{m,p}(\rho) \Phi_m(\varphi) \tag{A2.7}$$

The resonant dependence on wavelength in Eq. (A2.7) is determined by $G_{m,p}(\lambda,z,z_1)$, while $C_{m,p}$, $\Xi_{m,p}(\rho)$, and $\Phi_m(\varphi)$ are weak functions of wavelength that can be ignored. Eq. (A2.7) validates Eq. (2) with $C=C_{m,p}$ and $G(\lambda,z,z_1)=G_{m,p}(\lambda,z,z_1)$.

The transmission amplitude through the MF is found as [18]:

$$T(\lambda,z_1)=1-i\sum_{\mathbf{c}} \frac{|V_{\mathbf{c},\mathbf{k}}|^2}{E_{\mathbf{k}}-E_{\mathbf{c}}+i\Gamma_{\mathbf{c}}} \tag{A2.8}$$

Similar to the derivation of Eqs. (A2.6) and (A2.7), this equation is transformed into

$$T(\lambda,z_1)=1-i\sum_{\mathbf{c}} |C_{m,p}|^2 G_{m,p}(\lambda,z,z_1) \tag{A2.9}$$

and in the neighborhood of a resonance, $\lambda \approx \lambda_{mp}^{(0),\pm}$, to

$$T(\lambda,z_1)\big|_{\lambda \approx \lambda_{mp}} \approx 1-i|C_{m,p}|^2 G_{m,p}(\lambda,z,z_1), \tag{A2.10}$$

which coincides with Eq. (3) for $C=C_{m,p}$ and $G(\lambda,z,z_1)=G_{m,p}(\lambda,z,z_1)$.

### 3. The WGM distribution and transmission amplitude in the semiclassical approximation

*3.1. No turning points. Localization in a uniform SNAP device.*

Solving the Schrödinger equation, Eq. (1), in the semiclassical approximation [22] in the absence of turning points, we find the Green function,

$$G(\lambda,z_1,z_2)=\frac{1}{2i\beta^{1/2}(\lambda,z_1)\beta^{1/2}(\lambda,z_2)}\exp\left(i\int_{z_<}^{z_>}\beta(\lambda,z)dz\right), \quad z_<=\min(z_1,z_2),\ z_>=\max(z_1,z_2), \tag{A3.1}$$

and the WGM field distribution is found from Eq. (2). From Eq. (A3.1) and Eq. (3), the transmission amplitude through the MF is

$$T(\lambda)=1-\frac{|C|^2}{2\beta(\lambda,z_1)} \tag{A3.2}$$

For a uniform SNAP fiber, $r(z)=r_0$, the propagation constant is independent of $z$,

$$\beta(\lambda,z)=\beta_0(\lambda)=\pi n(2/\lambda_{res})^{3/2}(\lambda_{res}+i\gamma_{res}-\lambda)^{1/2} \tag{A3.3}$$

Substitution of this equation into Eq. (A3.1) shows that, as expected, the WGM field $\Lambda(\lambda,z_1,z)=CG(\lambda,z_1,z)$ exponentially vanish at both sides of the MF due to the presence of loss defined by $\gamma_{res}$. In the absence of attenuation, $\gamma_{res}=0$, in accordance with illustration in Fig. 2(b), the WGM amplitude $|\Lambda(\lambda,z_1,z)|$ is delocalized and is uniform along the axis $z$ for $\lambda<\lambda_{res}$, i.e., above the potential barrier. It is localized and exponentially decaying for $\lambda>\lambda_{res}$, i.e., in the underbarrier region. From Eqs. (A3.1), (A3.2), and (A3.3) the following expressions for the field amplitude $\Lambda(\lambda,z_1,z)$ and transmission amplitude $T(\lambda,z_1)$ are found:

$$\Lambda(\lambda,z_1,z)=\frac{C}{2i\beta_0(\lambda)}\exp(i\beta_0(\lambda)|z-z_1|), \tag{A3.2}$$

$$T(\lambda,z_1)=T_0(\lambda)=1-\frac{|C|^2}{2\beta_0(\lambda)}. \tag{A3.4}$$

The tree-dimensional problem of resonant transmission through a uniform SNAP device coupled to a MF was solved in [20] by calculation of the sum over the turns of circulating beam excited in the SNAP fiber. Eq. (A3.2) coincides with the expression for the WGM field amplitude obtained in [20]. However, Eq. (A3.4) corrects the expression for the transmission amplitude of Ref. [20], which mistakenly contained the factor $i$ in front of $|C|^2$. With this correction, the shape of the transmission resonance appears to be asymmetric rather than symmetric. In particular, for the small MF/SNAP fiber coupling $C$, the transmission power is

$$P(\lambda,z_1)=|T_0(\lambda)|^2=1-|C|^2\operatorname{Re}(\beta_0(\lambda)^{-1})$$
$$=1-\frac{|C|^2\lambda_{res}^{3/2}\{[(\lambda-\lambda_{res})^2+\gamma_{res}^2]^{1/2}+\lambda_{res}-\lambda\}^{1/2}}{4\pi n_f[(\lambda-\lambda_{res})^2+\gamma_{res}^2]^{1/2}} \tag{A3.5}$$

Fig. A1 compares the universal asymmetric shape of the resonant transmission power for the microcylinder, $P(\lambda,z_1)$, as a function of dimensionless wavelength $(\lambda-\lambda_{res})/2\gamma_{res}$ with the universal Lorentzian shape of a resonant transmission power $P_s(\lambda,z_1)$ through the fully localized state (e.g., for a microsphere resonator) with the same loss $\gamma_{res}$,

$$P_s(\lambda,z_1)=1-\frac{|C_s|^2}{(\lambda-\lambda_{res})^2+\gamma_{res}^2}. \tag{A3.6}$$

The ratio of the Q-factors of these resonators, which is inverse proportional to their FWHMs, is equal to 3.388.

*3.2. One turning point. Localization enforced by a point contact.*

If the radius variation of an optical fiber is monotonic, the presence of a MF may still leads to full localization of a WGM as illustrated in Fig. 2(c). In this case, solution of Eq. (1) exponentially vanishes to the left hand side of the turning point, $z<z_t$, and is an outgoing wave to the right hand side of the MF position, $z>z_1$. The Green's function of Eq. (1) found in the semiclassical approximation with these boundary conditions is

$$G(\lambda,z_1,z)=\frac{C\exp\left(-i\varphi(\lambda,z_t,z_1)-\frac{3\pi i}{4}\right)}{\beta^{1/2}(\lambda,z_1)\beta^{1/2}(\lambda,z)}\begin{cases}\frac{1}{2}\exp(-|\varphi(\lambda,z_t,z)|), & z<z_t,\\ \cos\left(\varphi(\lambda,z_t,z)-\frac{\pi}{4}\right), & z_t<z<z_1,\\ \frac{1}{2}\cos\left(\varphi(\lambda,z_t,z_1)-\frac{\pi}{4}\right)\exp(i\varphi(\lambda,z_1,z)), & z>z_1,\end{cases} \quad (A3.7)$$

$$\varphi(\lambda,z_1,z)=\int_{z_1}^{z}\beta(\lambda,z)dz. \quad (A3.8)$$

From this equation, light is confined between the turning point $z_t$ and the MF position $z_1$ if the following condition of destructive interference at $z_1$ is fulfilled:

$$\int_{z_t}^{z_1}\beta(\lambda,z)dz=\frac{3\pi}{4}+\pi n \quad (A3.9)$$

where $n$ is a large positive integer. Then, in the considered approximation, the transmission amplitude defined by Eq. (3) is equal to unity, i.e., light does not couple into the SNAP fiber and the localized state becomes dark.

*3.3. Two turning points. A bottle microresonator. Dark states.*

If the MF is located inside the bottle microresonator, which contains multiple localized states as illustrated in Fig 2(a), and is separated from the turning points by the distance greater than the axial wavelength $2\pi/\beta(\lambda,z)$, then Eq. (1) can be solved in the semiclassical approximation. In particular, the transmission amplitude, Eq. (3), is

$$T(\lambda,z_1)=1-i|C|^2\frac{\cos(\varphi(\lambda,z_{t1},z_1)+\frac{\pi}{4})\cos(\varphi(\lambda,z_1,z_{t2})+\frac{\pi}{4})}{2\beta(\lambda,z_1)\cos(\varphi(\lambda,z_{t1},z_{t2}))}, \quad (A3.10)$$

with $\varphi(\lambda,z_1,z_2)$ defined by Eq. (A3.8). The resonant peaks of the transmission amplitude correspond to the zeros of denominator in this equation $\lambda=\lambda_n+i\gamma_{res}$, which are determined by the quantization rule

$$\int_{z_{t1}}^{z_{t2}}\beta(\lambda,z)dz=\frac{\pi}{2}+\pi n. \quad (A3.11)$$

Expansion of Eq. (A3.10) near $\lambda_n+i\gamma_{res}$ yields:

$$T(\lambda,z_1)=1-\frac{i|C|^2\cos^2(\varphi(\lambda,z_{t1},z_1)+\frac{\pi}{4})}{2\beta(\lambda,z_1)\chi(\lambda,z_1)[(\lambda-\lambda_n)-i\gamma_{res}]},$$

$$\chi(\lambda,z_1)=\int_{z_{t1}}^{z_{t2}}\left.\frac{d\beta(\lambda,z_1)}{d\lambda}\right|_{\lambda=\lambda_n}dz.$$

(A3.12)

If the numerator in Eq. (A3.12) vanishes then the bottle microresonator becomes a dark state uncoupled from the MF.

It is instructive to compare the expression for the resonant transmission amplitude through a localized state, Eq. (A3.12), with the known formula:

$$T(\lambda,z_1)=\frac{(\lambda-\lambda_n)-i(\gamma_p-\gamma_c)}{(\lambda-\lambda_n)-i(\gamma_p+\gamma_c)},$$

(A3.13)

where parameters $\gamma_p$ and $\gamma_c$ determine the propagation loss of the SNAP fiber and the coupling to the MF, respectively. From Eqs. (A3.12) and (A3.13) we find $\gamma_{res}=\gamma_p+\gamma_c$ and

$$\gamma_c=-\frac{|C|^2\cos^2(\varphi(\lambda,z_{t1},z_1)+\frac{\pi}{4})}{4\beta(\lambda,z_1)\chi(\lambda,z_1)}$$

(A3.14)

### 4. Exact solution for the quadratic variation of the SNAP fiber radius

The semiclassical approximation of Section 3 fails near turning points and also near the bottom of quantum wells and the top of potential barriers. Solution of Eq. (1) in these regions is critical because it corresponds to the edge peaks of the resonant transmission amplitude of the bottle microresonators (Fig. 3) and also to largest principal peaks featuring slopes and neck regions of the fiber (Fig. 4). The radius variation in these cases can be approximated by the quadratic dependence:

$$r(z)=r_0+\frac{(z-z_0)^2}{2R},$$

(A4.1)

where the axial radius $R$ can be positive for the neck-shaped SNAP fiber and negative for the bottle-shaped fiber. Then, the Green function of Eq. (1) can be expressed through the parabolic cylinder functions. For numerical simulations, it is more convenient to use the integral representation [26]:

$$G(\lambda,z_1,z_2)=-\frac{\delta z_0}{2^{3/2}\pi^{1/2}}\int_0^\infty \frac{dx}{[\sinh(\sigma x)]^{1/2}}\exp\left\{\frac{i\pi}{4}-i\overline{\Delta\lambda}\sigma x+\frac{i}{2\sinh(\sigma x)}\left[\cosh(\sigma x)(\overline{z}_1^2+\overline{z}_2^2)-2\overline{z}_1\overline{z}_2\right]\right\},$$

(A4.2)

where $\sigma=(\text{sign}(R))^{1/2}$,

$$\overline{z}_j = z_j / \delta z_0 \text{ and } \overline{\Delta\lambda} = (\lambda - \lambda_{res}) / \delta\lambda_0 \tag{A4.3}$$

are the dimensionless axial coordinate and wavelength, and

$$\delta z_0 = (\lambda_{res}/2\pi n_{eff})^{1/2}(r_0 R)^{1/4} \text{ and } \delta\lambda_0 = \lambda_{res}^2 (r_0 R)^{-1/2}/2\pi n_{eff} \tag{A4.5}$$

are the characteristic lengths along $z$ and $\lambda$. Eq. (A4.2) is valid for $R > 0$.

## 5. Experimental measurement of the fiber radius variation

In the semiclassical approximation, the relation between the resonance amplitude $T(\lambda, z_1)$ and SNAP fiber radius variation $\Delta r(z)$ can be clarified with Fig. 2. If the MF is situated in the region where the SNAP fiber has a bottle shape (Fig. 2(a)), the spectra $T(\lambda, z_1)$ possess sharp resonances corresponding to localized WGMs (see, e.g., Fig. 3). For each MF position $z_1$, the resonant peaks of the amplitude are located in the interval between the turning point wavelength, $\lambda_{Turn}(z_1) = \lambda_{res}(1 + \Delta r(z_1)/r_0)$ and the wavelength $\lambda_{Top}$ corresponding to the top of the potential barrier and independent of $z_1$. Thus, the radius variation is found as

$$\Delta r(z_1) = (\lambda_{Turn}(z_1) - \lambda_{res})r_0 / \lambda_{res} \tag{A5.1}$$

As an example, the fiber radius variation, shown in Fig. 3(c), envelopes the resonance spectra of Fig. 3(b).

In the case of extremely slow variation of the fiber radius, reflection of WGMs from the turning points can be ignored. The peak of this amplitude $\lambda_{Peak}(z_1)$ is shifting proportionally to $\Delta r(z_1)$ so that

$$\Delta r(z_1) = (\lambda_{Peak}(z_1) - \lambda_{res})r_0 / \lambda_{res} \tag{A5.2}$$

Eq. (A5.2) resembles the approach of Refs. [24,25]. Finally, for a SNAP fiber segment, which does not contain localized bottle states, (e.g., the regions near the fiber neck (Fig. 2(b)) and to the right hand side of the neck (Fig. 2(c)), the resonance amplitude has the Airy-type oscillations (see, e.g., Fig. 4 (b)) It can be shown numerically, that, in this case, the fiber radius variation is proportional to the shift of the largest main peak of the resonance amplitude, $\lambda_{MainPeak}(z_1)$:

$$\Delta r(z_1) = (\lambda_{MainPeak}(z_1) - \lambda_{res})r_0 / \lambda_{res} \tag{A5.3}$$

Generally, the fiber radius variation can be found from the amplitude spectra $T(\lambda, z_1)$ measured at different MF positions $z_1$ by numerical solution of the inverse problem for Eqs. (1) and (3).

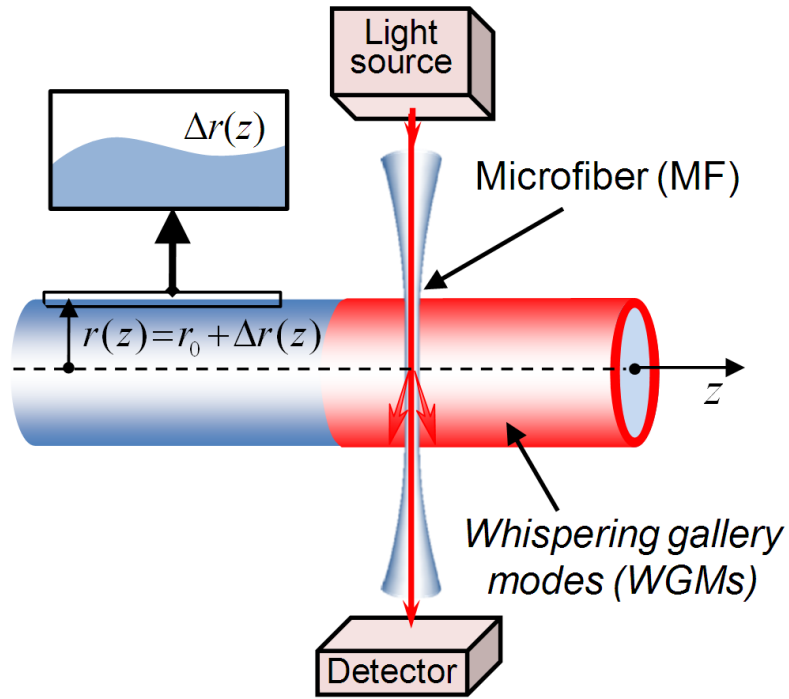

Fig. 1. Illustration of a SNAP device. The whispering gallery modes (WGMs) are excited in the SNAP fiber with a transverse microfiber (MF) connected to the light source and detector. The nanoscale radius variation of this fiber, $\Delta r(z)$, determines the distribution of slow WGMs along the fiber surface and also the spectrum of light transmitted through the MF.

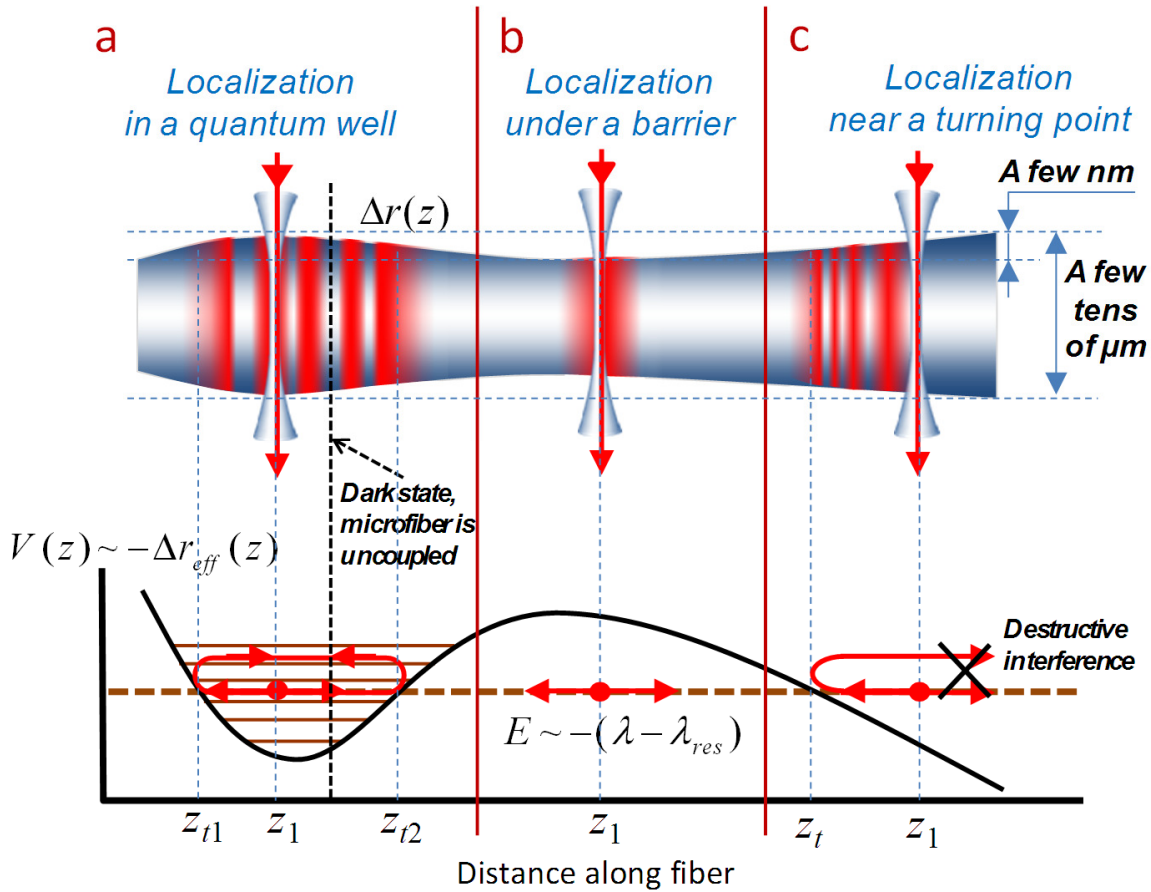

Fig. 2. Basic elements of a SNAP device. (a) – MF coupling to a bottle microresonator. Slow WGMs excited by the MF can be trapped in the quantum well formed by the nanoscale variation of the effective fiber radius. (b) – MF coupling to the neck of the fiber. In this case, variation of the fiber radius forms a potential barrier. (c) – MF in the region of monotonic variation of the fiber radius. In this case, the WGMs emitted by the MF are reflected from the turning point and, under certain conditions, can be completely halted by the MF source.

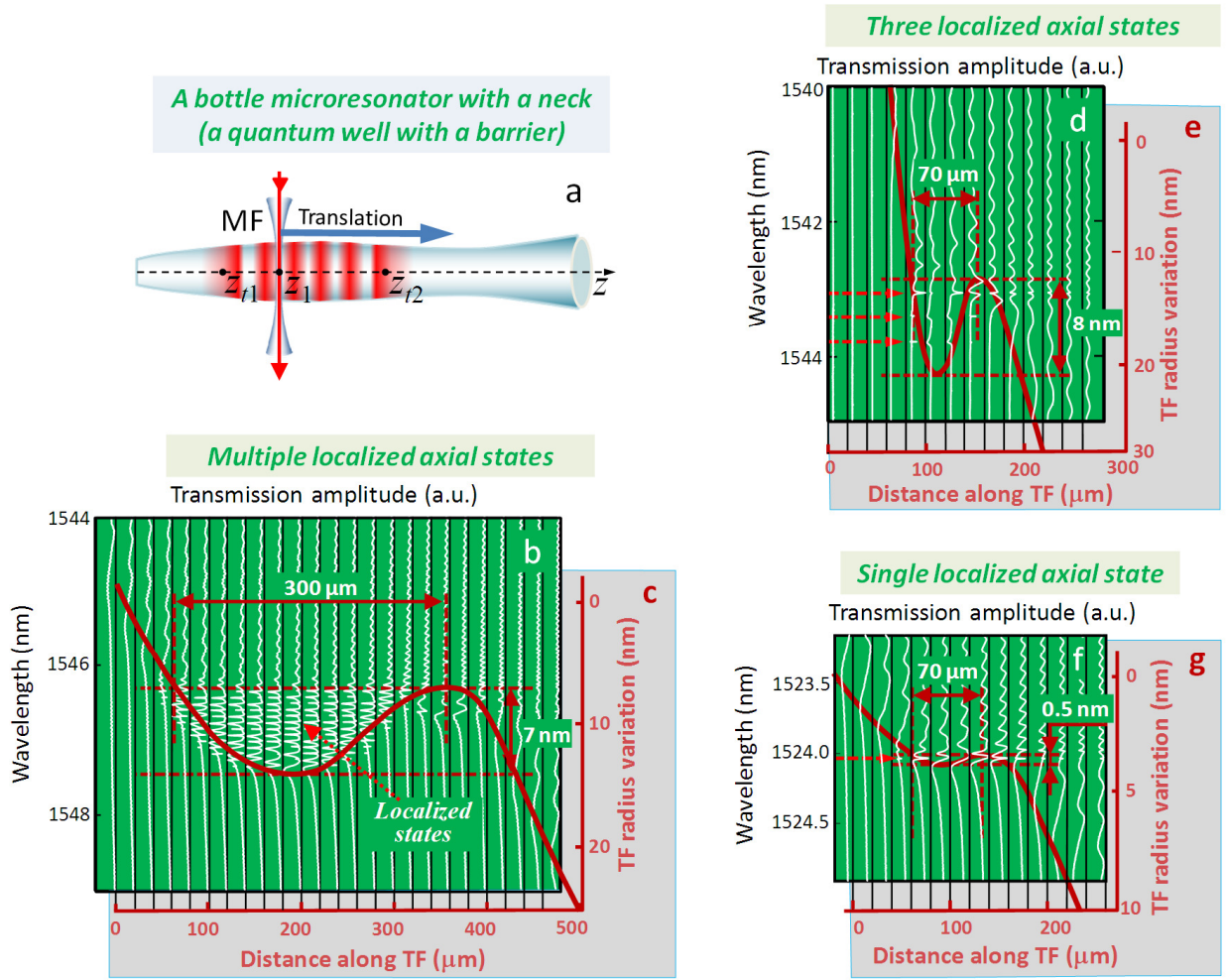

Fig. 3. Experimental characterization of the nanoscale radius variation and spectra of the fabricated bottle microresonators with a neck. (a) – Illustration of a bottle microresonator with a neck; (b), (d), and (f) – Resonant transmission spectra measured for the MF positions along the SNAP fiber spaced by 20 μm for the microresonator with multiple, three, and one axial state, respectively; (c), (e), and (g) – radius variation of the SNAP fiber in cases (b), (d), and (f).

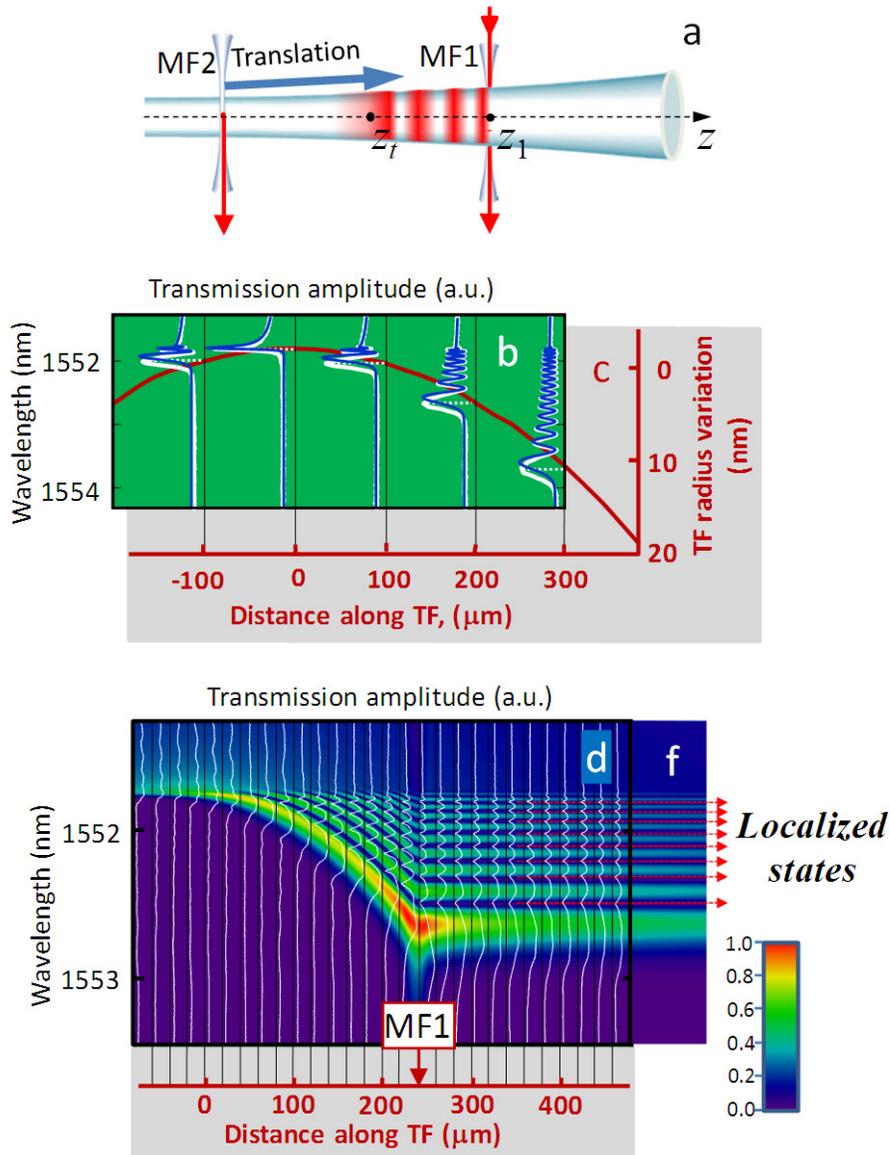

Fig. 4. Experimental demonstration of the effect of full localization of light between a turning point and a point contact of a SNAP fiber with an MF. (a) – Illustration of the localized WGM field distribution excited by MF1 and scanned by MF2. (b) – Experimental resonance spectra along the SNAP fiber for the MF positions spaced by 100 μm (white curves) compared with theoretically found resonance spectra (blue curves). (c) – Radius variation of the SNAP fiber found from spectra in Fig. 4(b); (d) – Resonant transmission spectra measured for the MF2 positions along the SNAP fiber spaced by 20 μm and MF1 positioned at 240 μm from the minimum of the fiber radius; (f) – Surface plot of the theoretically calculated WGM field distribution as a function of the distance along the SNAP fiber and wavelength for the experimental radius variation found from Fig. 4(c).

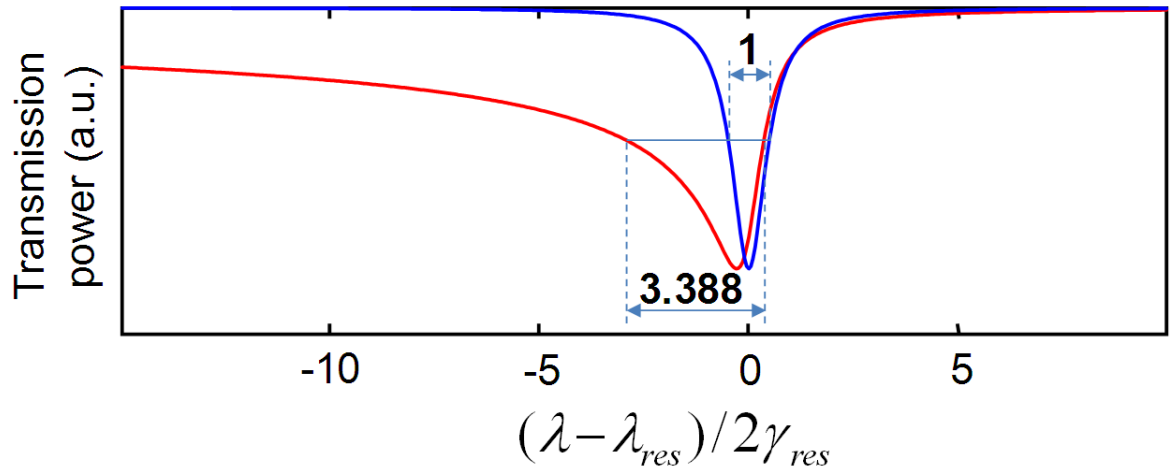

Fig. A1. The shapes of the transmission power of the cylindrical resonance (red curve) compared to the resonance at a fully localized state (blue curve).